\documentclass[%
 reprint,
 amsmath,amssymb,
 aps,
]{revtex4-2}

\usepackage{graphicx}
\usepackage{dcolumn}
\usepackage{bm}
\usepackage{physics}
\usepackage{mathtools}
\usepackage{float}
\usepackage[usenames,dvipsnames]{color}

\begin{document}

\preprint{APS/123-QED}

\title{Tuning the probability detection of OAM entangled photons in \\ Helical Ince-Gauss modes}


\author{A.A. Aguilar-Cardoso}
\email{aaron.cardoso.aa@gmail.com}

\author{D. Rodríguez-Guillen}%
 

\author{R. Ramírez-Alarcón}
\email{roberto.ramirez@cio.mx}

\affiliation{
Centro de Investigaciones en Óptica A. C., Loma del Bosque 115, Colonia Lomas del Campestre, 37150 León Guanajuato, México
}%

\date{\today}

\begin{abstract}

In this work we provide a detailed theoretical and experimental analysis of the two-photon Orbital Angular Momentum (OAM) entangled state, generated by a type-I Spontaneous Parametric Down Conversion (SPDC) process, when decomposed in terms of the Helical Ince-Gauss (HIG) modes basis. By exploiting the unique characteristics of this modal basis we show how the probability detection of the photon-pair can be tuned with the ellipticity parameter of the modes. We also show that, on the HIG basis the SPDC state has the contribution of two different symmetric Bell states, and it is possible to maximize the probability of each HIG symmetric Bell state separately, also by tuning the elipticity of the projected basis. The observed properties are confirmed experimentally by implementing measurements of the HIG modal joint probability of the SPDC two-photon state and Bell-type inequality violation experiments. 
\end{abstract}

\maketitle


\section{Introduction}
The study of entangled photon pairs has expanded beyond traditional degrees of freedom like spin and polarization, leading to the exploration of alternative entanglement degrees of freedom, such as energy-time entanglement~\cite{shalm-2012,cuevas-2013}, momentum-position entanglement~\cite{howell-2004}, and frequency-transverse momentum entanglement~\cite{zielnicki-2018}, but photons pairs can also be entangled in the Orbital Angular Momentum (OAM) degree of freedom~\cite{mair-2001,leach-2009}.

In general, a two-photon OAM entangled state is generated through parametric processes like Spontaneous Parametric Down-Conversion (SPDC) in nonlinear crystals, where a high-energy photon is annihilated to produce two lower-energy photons, when satisfying phase matching conditions. The process of type-I SPDC in its collinear configuration preserves OAM when all the emitted wavevectors are collected in the experimental setup~\cite{torres-2003,osorio-2008,ibarra-borja-2019}, then under these conditions the spatial structure of the emitted photons can be expressed as a coherent superposition of products of Laguerre-Gauss (LG) Fock states, which satisfy OAM conservation ~\cite{allen-1992, van-enk-1994, calvo-2006}. 

Several detailed studies have been carried out of the two-photon OAM entangled state projected on different modal basis, such as Hermite-Gauss (HG) modes~\cite{walborn-2005}, Bessel-Gauss (BG) modes~\cite{mclaren-2012}, or Hermite-Laguerre-Gaussian (HLG) modes~\cite{tang-2016}. Various applications have also been developed from this entangled state described in different modal basis~\cite{nape-2018,wang-2020,zhu-2020}. Another possibility to study OAM entangled two-photon states is to use the Helical Ince-Gauss (HIG) modes~\cite{bandres-2004b}. 

The Ince-Gauss (IG) modes are an exact, complete, and orthogonal solution to the paraxial wave equation solved in elliptical coordinates. Its transverse mode structure is defined by the discrete order $p$ and degree $m$ indexes of the Ince polynomials, and also by the ellipticity $\epsilon$, which is a continuous and dimensionless parameter that modulates the elliptical shape of the transverse structure of the beam. A superposition of the IG modes gives place to the Helical Ince-Gauss (HIG) modes basis, which main characteristic is that it represents the transition basis between the LG modes and the Helical HG modes. The HIG modes have unique properties, for instance, it has been proved that it is possible to tune the OAM content of the modes~\cite{plick-2013} by varying the ellipticity parameter, making this basis and interesting scenario to study OAM entangled two-photon states. 

In this work we provide a detailed study of the two-photon OAM entangled state, produced by a type-I collinear SPDC process when decomposed in terms of the HIG modes basis. By doing so, we have found that the OAM entangled state can be written as a sum of two different symmetric Bell states, whose probability coefficients are dependent on the ellipticity that defines the HIG basis. This dependence of the probability coefficients with the ellipticity of the basis confirms that it is possible to tune the modal probability of the state. Additionally, we demonstrate that it is possible to maximize the probabily of each HIG symmetric Bell state separately, by adjusting the ellipticity of the selected basis. We tested our predictions by performing measurements of HIG modal joint probability of the SPDC two-photon state and Bell-type inequality violation experiments. We believe our results could be useful for quantum communication protocols with high dimensional OAM entangled states.

\section{Ince-Gauss modes}
We start with the paraxial wave equation (PWE), given by,
\begin{equation} \label{PWE}
    \nabla_t^2 u(\vec{r})+2ik\frac{\partial}{\partial z}u(\vec{r})=0,
\end{equation}

\noindent where the transverse Laplacian is $\nabla_t^2=\frac{\partial^2}{\partial x^2}+\frac{\partial^2}{\partial y^2}$, $k$ is the wave number, and $u(\vec{r})$ represents the complex amplitude of the optical wave along the propagation direction $z$.

\begin{figure}[h]
\includegraphics[scale=0.35]{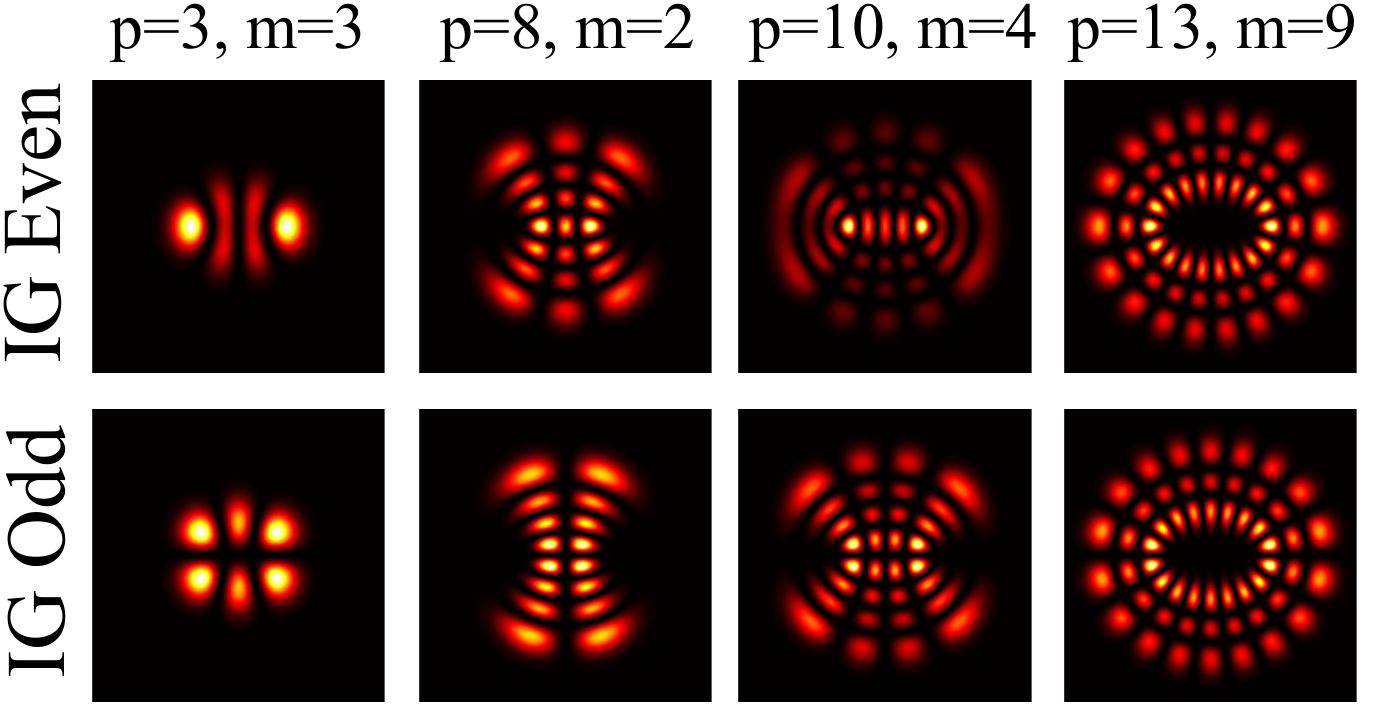}
\caption{\label{modosIGeo}Transverse intensity distributions of even and odd Ince-Gauss modes with $\epsilon=3$ and various $p$ and $m$ values.}
\end{figure}

By solving Eq.~(\ref{PWE}) in different coordinate systems we obtain different families of solutions for instance, LG and HG modes are families of solutions to the PWE in cylindrical $(r,\phi,z)$ and Cartesian $(x,y,z)$ coordinates, respectively. In our case we are interested in the solutions to the PWE in elliptical coordinates $(\xi,\eta,z)$, namely the even and odd Ince-Gauss modes (z=0)~\cite{bandres-2004b},
\begin{equation}
 \textrm{IG}^e_{p,m}(\Vec{r},\epsilon)=\textit{C}C_p^m(i\xi,\epsilon)C_p^m(\eta,\epsilon) e^{\frac{-r^2}{\omega^2(0)}},
\end{equation}
\begin{equation}
\textrm{IG}^o_{p,m}(\Vec{r},\epsilon)=\textit{S}S_p^m(i\xi,\epsilon)S_p^m(\eta,\epsilon)e^{\frac{-r^2}{\omega^2(0)}},
\end{equation}

\noindent where $C_p^m$ and $S_p^m$ are solutions of the Ince differential equation known as the even and odd Ince polynomials respectively, of order $p$ and degree $m$. The modal parameters $p$ and $m$ are integer numbers, always have the same parity and meet $0 < m < p$ for even functions and $1 < m < p$ for odd functions. The ellipticity $\epsilon$ is a positive and continuous parameter, which determines the elliptical structure of the beam. It is defined as $\epsilon=2f_0^2/\omega_0^2$, where $f_0$ and $\omega_0$ are physical scale parameters of the mode at $z=0$, known as the semifocal separation and the beam width respectively. The transverse profile of an Ince-Gauss mode can take on a variety of shapes as seen in Fig.~\ref{modosIGeo}.

\begin{figure}[H]
\includegraphics[scale=0.35]{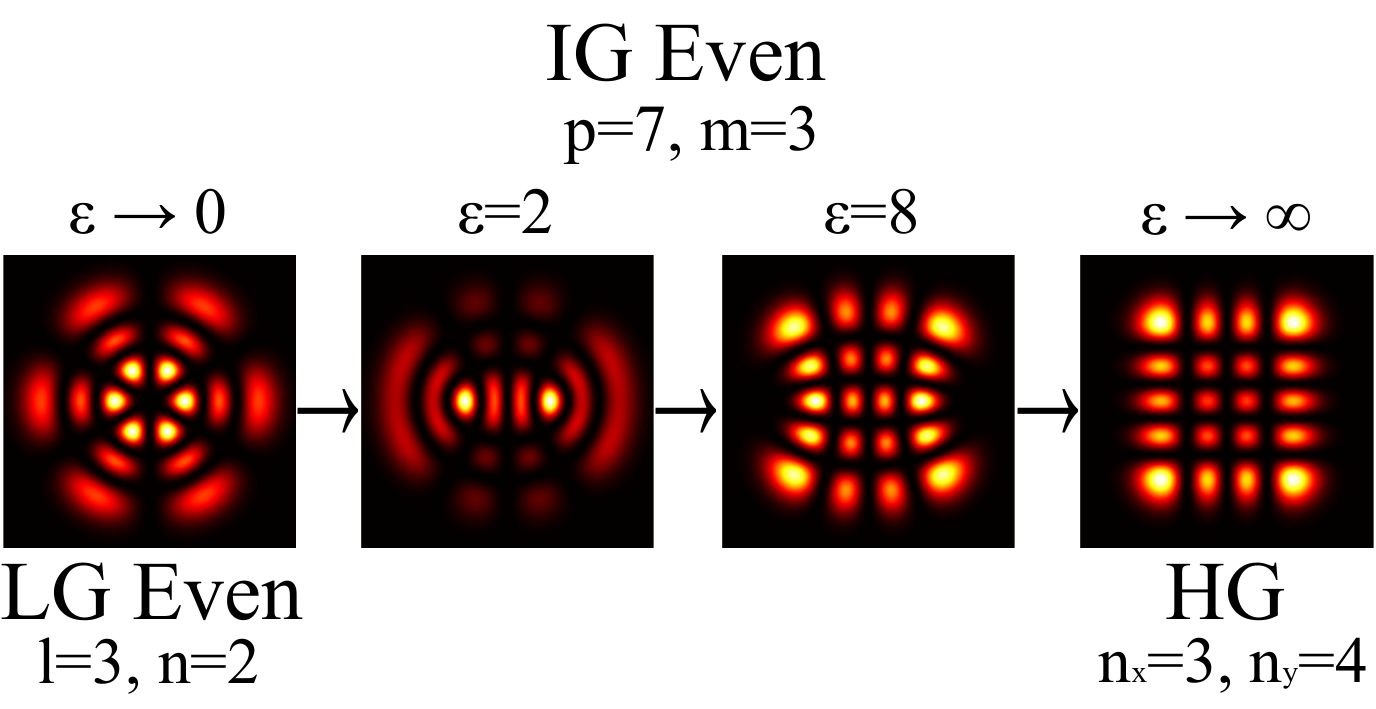}
\caption{\label{trans} By varying the ellipticity of the IG modes we can make the transition from the even and odd LG to the HG modes. As an example, when $\epsilon\to0$, the mode $\textrm{IG}^e_{7,3}$ tend to $\textrm{LG}^e_{3,2}$, and when $\epsilon\to\infty$, the mode $\textrm{IG}^e_{7,3}$ tend to $\textrm{HG}_{3,4}$.}
\end{figure}

The ellipticity of the Ince-Gauss modes is quite a unique parameter, it can take continuous values, contrary to $p$ and $m$, and even more, since the elliptic coordinates can approximate to cylindrical and Cartesian coordinates, the relation of Ince-Gauss modes with the Hermite and Laguerre Gaussian modes is given via the ellipticity; larger absolute values result in a more pronounced elliptical profile, while smaller values indicate a closer approximation to a circular shape. So, in the limit $\epsilon\to 0$ the elliptic coordinates tend to the circular cylindrical coordinates and the even and odd Ince-Gauss modes tend to the even and odd Laguerre-Gauss modes, with the indices relations: $m=l$ and $p=2n+l$. On the contrary limit, when $\epsilon\to \infty$ the elliptic coordinates tend to the Cartesian coordinates, with the even and odd Ince-Gauss modes tending to Hermite-Gauss modes with the indices relations $n_x=m$ and $n_y=p-m$ for even Ince-Gauss, and $n_x=m-1$ and $n_y=p-m +1$ for odd Ince-Gauss. A example of this transition is shown in Fig.~\ref{trans} for the even IG mode $p=7$, $m=3$.

As is the case for the LG and HG families of modes, the IG modes satisfy an orthonormality condition, given by
\begin{equation}\label{orto}
    \int\int_{-\infty}^\infty \textrm{IG}^\sigma_{p,m}(\Vec{r},\epsilon) \textrm{IG}^{\sigma',*}_{p',m'}(\Vec{r},\epsilon') \textbf{dS} =\delta_{\sigma \sigma'}\delta_{p p'}\delta_{m m'},
\end{equation}

\noindent where $\sigma$ is the parity of the mode (e,o). Importantly, note that ellipticity parameter is not present in the orthonormality condition of the IG modes.

Now, as these modes are complete families of solutions to the PWE, we are able to express any IG mode as a finite superposition of either LG modes or HG modes. However, we are only interested in the former case, expressed as
\begin{equation}\label{superpostion}
 \textrm{IG}^\sigma_{p,m}(\vec{r},\epsilon)=\sum_{l,n}D_{l n}^{\sigma} (\epsilon)\textrm{LG}_{l,n}^\sigma (\vec{r}),
\end{equation}

\noindent where $D_{ln}^\sigma(\epsilon)$ are the weights of the Laguerre-Gauss expansion. The LG modes that make up the superposition of a specific IG mode must meet the constraint $p=2n+l$, so the number of terms in Eq.~(\ref{superpostion}) is given by the value of $p$, having a total of $p -\lceil \frac{p}{2}\rceil + 1$ terms. These coefficients meet the condition $\sum_{l,n}D_{l n}^{\sigma 2} (\epsilon)=1$ and depends explicitly on the ellipticity and the parity, therefore by changing the ellipticity of the IG modes the expansion coefficients are also changing, giving more or less weight to each LG component. The $D_{ln}^\sigma(\epsilon)$ coefficients can be obtained as in Ref.~\cite{bandres-2004b}.

An important property of LG modes is that they carry integer values of OAM, due to its azimuthal angular dependence $e^{il\phi}$~\cite{allen-1992,barnett-2017,andrews-2012}, while the HG and IG beams do not have such angular dependence. In order to build IG modes with a rotating phase we must define the Helical Ince-Gauss (HIG) modes as a superposition of even and odd IG, given by
\begin{equation} \label{HIG}
\textrm{HIG}_{p,m}^\pm(\vec{r},\epsilon)=\frac{1}{\sqrt{2}}\left(\textrm{IG}_{p,m}^e(\vec{r},\epsilon)\pm i\textrm{IG}_{p,m}^o(\vec{r},\epsilon)\right).
\end{equation}

\begin{figure}[h]
\includegraphics[scale=0.35]{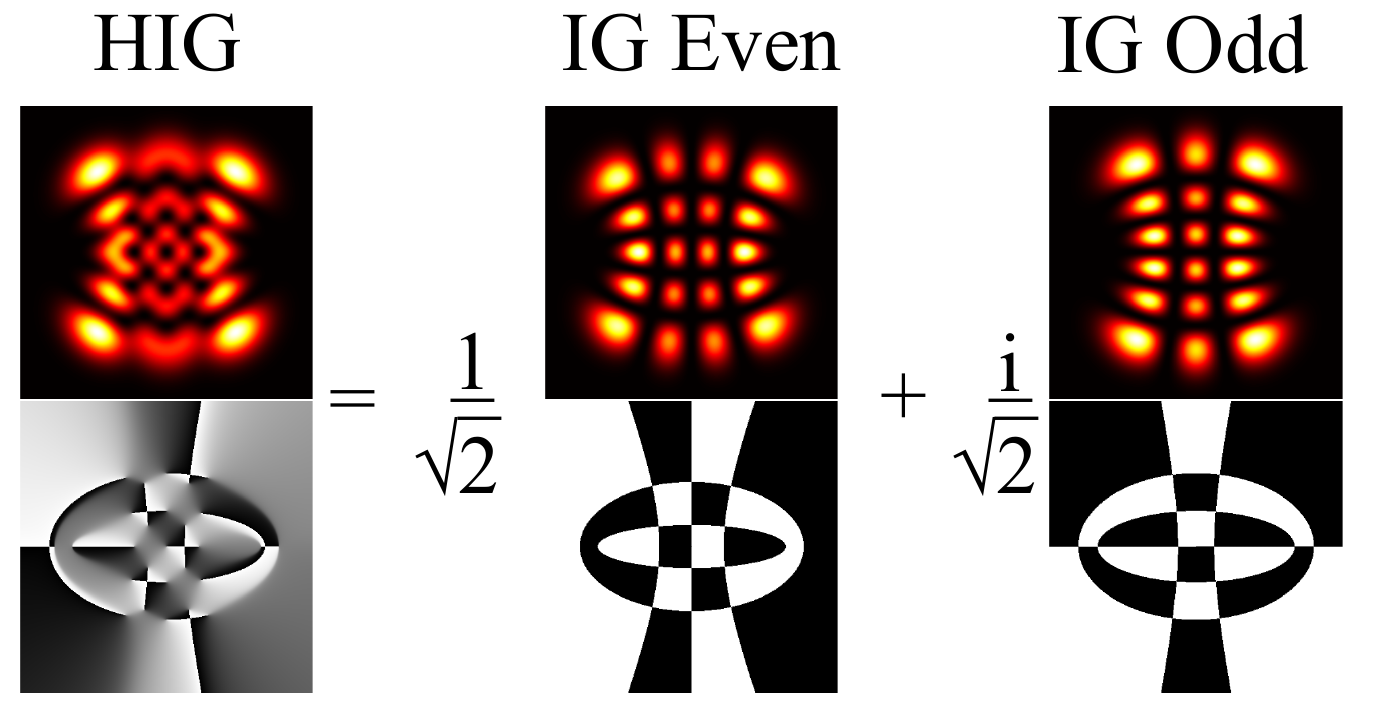}
\caption{\label{HIGmode} Superposition of even and odd IG modes gives the HIG modes. As an example $\textrm{HIG}_{7,3}^\pm(\vec{r},8)=\frac{1}{\sqrt{2}}\left(\textrm{IG}_{7,3}^e(\vec{r},8)+ i\textrm{IG}_{7,3}^o(\vec{r},8)\right)$ is shown. The phase rotates counterclockwise, as is the positive case of the HIG mode.}
\end{figure}

For the HIG, the number of elliptical rings is given by $1+(p - m)/2$, and each ring splits in single nodes as the ellipticity increase, as shown in Fig.~\ref{HIGmode}. The sign $\pm$ of the spatial function defines the phase rotating direction, where $+$ sign rotate the phase counterclockwise, and the $-$ sign rotate the phase clockwise. For the HIG modes the value of $m=0$ is not allowed, since the odd mode $IG_{p,m}^o$ is not defined for this case. 

Similar to the even and odd IG modes, as $\epsilon\to0$, the HIGs modes transform into corresponding LG modes, with the central singularities of the modes shifting towards the center of the beam. On the other hand, when $\epsilon\to\infty$, the HIG modes take the form of Helical Hermite-Gauss modes.

\section{SPDC two-photon OAM Entangled State in the HIG basis}
\label{section2}

The formalism used so far has been classical, however it is possible to study the paraxial electromagnetic field with quantum field theory. The transverse electromagnetic field is quantized by expanding the fields in any complete set of transverse vector modal functions that represents the state of the radiation field~\cite{calvo-2006}. So we can represent a single-photon number state created in a LG mode with $l$ of sign $\pm$ and $n$ modal numbers as a LG Fock state $\ket{L_ {l,n}^\pm}$. These states are eigenvectors to the paraxial OAM operator $\hat{L}_{z}$, also derived based on the paraxial approximation, with eigenvalue $\pm \hbar l$~\cite{allen-1992,barnett-2017}. So the OAM content of the LG Fock states is $\bra{L_ {l,n}^\pm}\hat{L}_{z}\ket{L_ { l,n}^\pm}=\pm \hbar l$. We can also represent a single-photon number state created in a HIG mode with $p$, $m$ and $\epsilon$ modal numbers as a HIG Fock state $\ket{I^{\pm,\epsilon}_ {p,m}}$, these states fulfill exactly the same properties that were described in the previous section. We can calculate the OAM content of these states with $\hat{L}_{z}$~\cite{plick-2013},
\begin{equation} \label{OAMfraccionario}
    \bra{I^{\pm,\epsilon}_ {p,m}}\hat{L}_{z}\ket{I^{\pm,\epsilon}_ {p,m}}=\pm\sum_{l,n}\hbar lD_{ln}^e(\epsilon)D_{ln}^o(\epsilon).
\end{equation} 

The main feature of this quantity is that there is no integer part of OAM, giving place to fractional expectation values of OAM per photon, which changes with the ellipticity parameter. The OAM of the HIG modes for $\epsilon\to0$ is equal to the value of $m$, since indices relations to the LG modes are $m = l$ and $p = 2n + l$, therefore, in this limit there are different HIG states for which the OAM value is exactly the same, but as the ellipticity increases, these values become detached from each other. 


A type-I collinear SPDC process generates photon pairs entangled in their OAM degree of freedom~\cite{mair-2001,andrews-2012,leach-2009}. If we focus on the case where the nonlinear crystal is pumped with a Gaussian beam ($l_p=0$) and the experimental setup allows for the collection of all of the emitted wavevectors then, the conservation law for the OAM of the signal and the idler photons $l_s=-l_i\equiv l$ is fulfilled. With this conditions, the entangled state generated by the SPDC process can be decomposed into a coherent superposition of LG Fock states, which satisfy OAM conservation, as given by

\begin{gather}
\ket{\Psi_{\textrm{SPDC}}}=\nonumber\\
\sum_{l,n_s,n_i}C_{n_s,n_i}^{l,-l} \left( \ket{L_{l,n_s}^+}_s\ket{L_{l,n_i}^-}_i +\ket{L_{l,n_s}^-}_s\ket{L_{l,n_i}^+}_i\right),
\label{estadoSPDC}
\end{gather}

\noindent where the coefficients $|C_{n_s,n_i}^{l,-l}|^2$ represents the joint probability of finding one photon in the LG Fock state with positive helicity and the other with negative helicity~\cite{miatto-2011}. Such state represents an increase in the dimension of the Hilbert space, in contrast with polarization entangled states, as it can be as large as $l$ indices are taken~\cite{krenn-2017}.


Now, in this work we seek to describe two-photon OAM entangled state in terms of HIG Fock states, as these constitute a complete orthonormal basis. To do so, we make a change of basis by projecting the SPDC state into the HIG Fock states with the same $p$ and $m$ parameters, but with distinct ellipticity for each photon. The latter is possible since the family of IG solutions is orthogonal with respect to the modal numbers and parity of the state, but not to ellipticity (Eq.~(\ref{orto})). To proceed with this calculation we take advantage of the fact that the HIG modes are a coherent and finite superposition of the LG modes (Eq.~(\ref{superpostion})), whose expansion coefficients are explicit functions of the ellipticity. Since it is not of our interest to study cross correlations of the HIG states, we are going to consider that cross correlations between eigenstates $n_s$ and $n_i$ are small enough to neglect them, leaving only $n_s=n_i\equiv n$ states. This is realized experimentally by reducing size of signal and idler beam, compared to the size of the pump beam~\cite{miatto-2011}. 

When performing the change of basis to HIG Fock states, we find that there is a probability of finding both down converted photons with the same helicity, which does not happen on the LG basis (Eq.~\ref{estadoSPDC}). Thereby, we can describe the SPDC state as a superposition of the symmetric HIG Bell-states $\ket{\Psi^+}$ and $\ket{\Phi^+}$, as

\begin{widetext}

\begin{equation}\label{SPDCIGdos}
\ket{\Psi_{\textrm{SPDC}}}=\sum_{p,m}\left[F_{p,m}\left(\ket{I^{+,\epsilon_s}_{p,m}}_s\ket{I^{-,\epsilon_i}_{p,m}}_i + \ket{I^{-,\epsilon_s}_{p,m}}_s\ket{I^{+,\epsilon_i}_{p,m}}_i\right)+G_{p,m}\left(\ket{I^{+,\epsilon_s}_{p,m}}_s\ket{I^{+,\epsilon_i}_{p,m}}_i + \ket{I^{-,\epsilon_s}_{p,m}}_s\ket{I^{-,\epsilon_i}_{p,m}}_i\right)\right],
\end{equation}
\end{widetext}

\noindent where expansion coefficients are given by
\begin{equation}\label{FCoef}
F_{p,m}=\frac{1}{2}\sum_{l,n}C_{n,n}^{l,-l}[D_{ln}^{e}(\epsilon_s)D_{ln}^{e}(\epsilon_i)+ D_{ln}^{o}(\epsilon_s)D_{ln}^{o}(\epsilon_i)],
\end{equation}
\begin{equation}\label{GCoef}
G_{p,m}=\frac{1}{2}\sum_{l,n}C_{n,n}^{l,-l}[D_{ln}^{e}(\epsilon_s)D_{ln}^{e}(\epsilon_i)- D_{ln}^{o}(\epsilon_s)D_{ln}^{o}(\epsilon_i)].
\end{equation}

\begin{figure}[h!]
\includegraphics[scale=0.35]{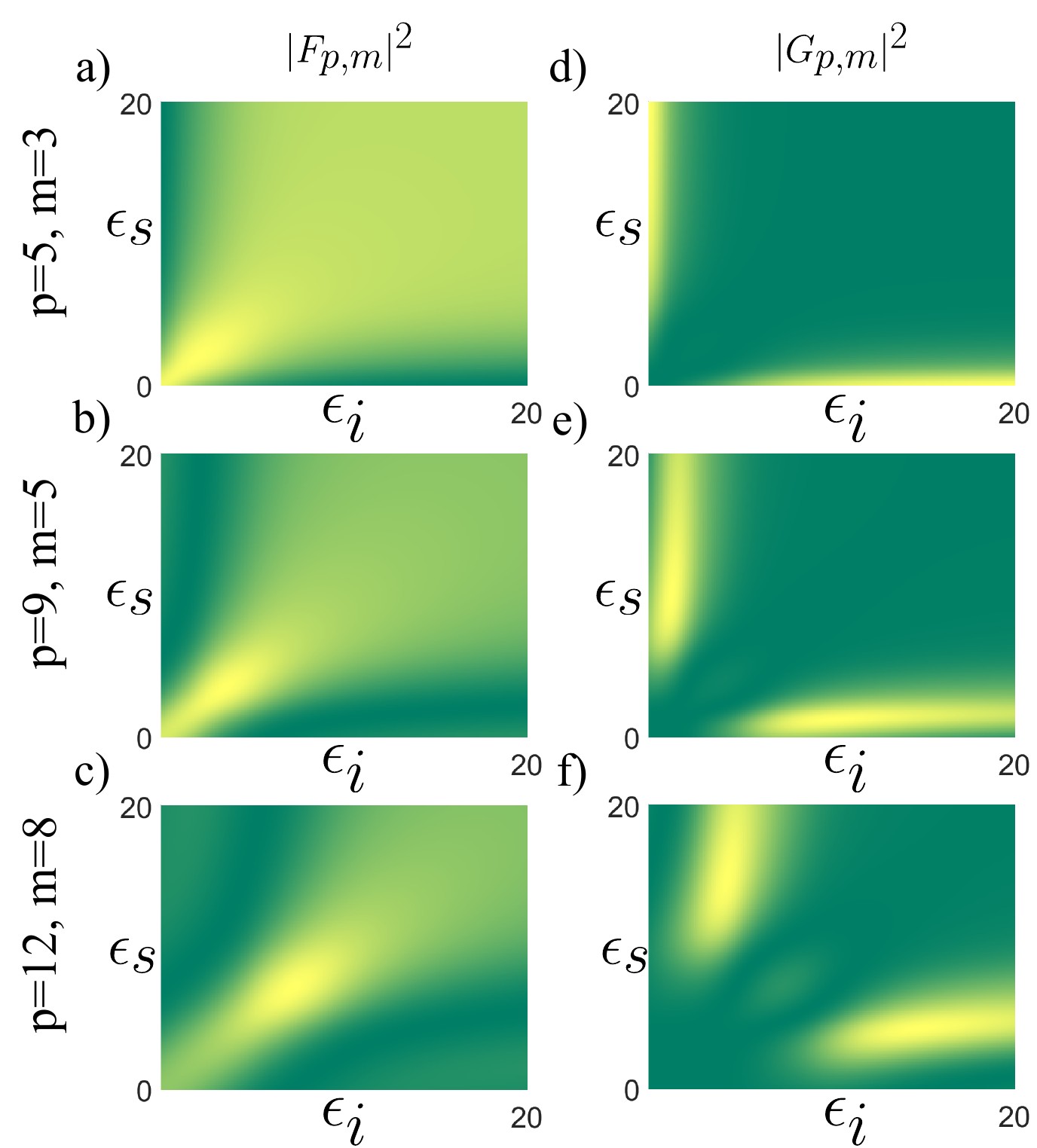}
\caption{\label{FIG:comparacion} Density plots of the $F_{p,m}$ and $G_{p,m}$ coefficients in terms of the photons ellipticity, for different HIG modes.}
\end{figure}

Both coefficients, $F_{p,m}$ and $G_{p,m}$, are truncated by the $D_{ln}^\sigma$ terms, such that $p=2n+l$, then, the number of terms is given by the value of $p$, having a total of $p -\lceil \frac{p}{2}\rceil + 1$ terms. It is also noteworthy that when $\epsilon_{s,i}\to 0$ then $G_{p,m}\to 0$, $F_{p,m}\to \sum_{l,n}C_{n,n}^{l,-l}$ and $\ket{I^{\pm,\epsilon}_{p,m}}\to\ket{L^\pm_{l,n}}$, and the photon-pair entangled state in the OAM basis of Eq.~(\ref{estadoSPDC}) is recovered. Figure~\ref{FIG:comparacion} shows the normalized squared absolute value of these coefficients for various $p$ and $m$ modal numbers. 

Equation~(\ref{SPDCIGdos}) represents the two-photon OAM entangled state generated by a collinear type-I SPDC process, described on the HIG basis. This description is our main result given all its relevant properties, as explained below. Firstly, it is noteworthy that the contribution of the $\ket{\Phi^+}$ state has never been taken into account~\cite{krenn-2013,zhu-2020} and even when the maximum value for $|F_{p,m}|^2$ is always greater than the maximum value of $|G_{p,m}|^2$, for any $p$ and $m$ values, the contribution of the $\ket{\Phi^+}$ state is not negligible at all since, as is shown in the next section, it is possible to measure the correlations generated by such state.

We can also notice that the coefficients in Eq.~(\ref{SPDCIGdos}) vary with the ellipticity, therefore, as we illustrate in the following sections, by tuning the ellipticity of the HIG modes, in which the photon-pairs are projected, it is possible to tune the probability of detection of each of the HIG Fock states.

Furthermore, the ellipticity values that maximize the probability of finding both photons with equal helicity ($|F_{p,m}|^2$) are the same values that minimize the opposite case ($|G_{p,m}|^2$), as seen in Fig.~\ref{FIG:comparacion}. Then, after post-selection (that means fixing p, m and $\epsilon_{s,i}$) it is possible to maximize the probability $|F_{p,m}|^2$ ($|G_{p,m}|^2$), thus selecting a $\ket{\Psi^+}$ ($\ket{\Phi^+}$) Bell state component.

In summary, after post-selection it is possible to tune the probability of finding the photon-pair state in a specific HIG Bell-state, by varying the ellipticity of the projected mode. Also, it is possible to maximize the contribution between $\ket{\Psi^+}$ or $\ket{\Phi^+}$, by finding which value of $\epsilon_s$ and $\epsilon_i$ maximize either $|F_{p,m}|^2$ or $|G_{p,m}|^2$, which depends on the desired case according to Eq~(\ref{FCoef}) and Eq~(\ref{GCoef}).  In this way, the unique ellipticity parameter and its effect on tunning the probability detection of HIG modes could find application in quantum communication protocols, making the HIG modes an interesting alternative to standard OAM basis such as LG modes. In the next section, the existence of such tuning and of both states is experimentally demonstrated.

\section{Experiments}
\label{experiments}

\begin{figure} [h!]
\includegraphics[scale=0.5]{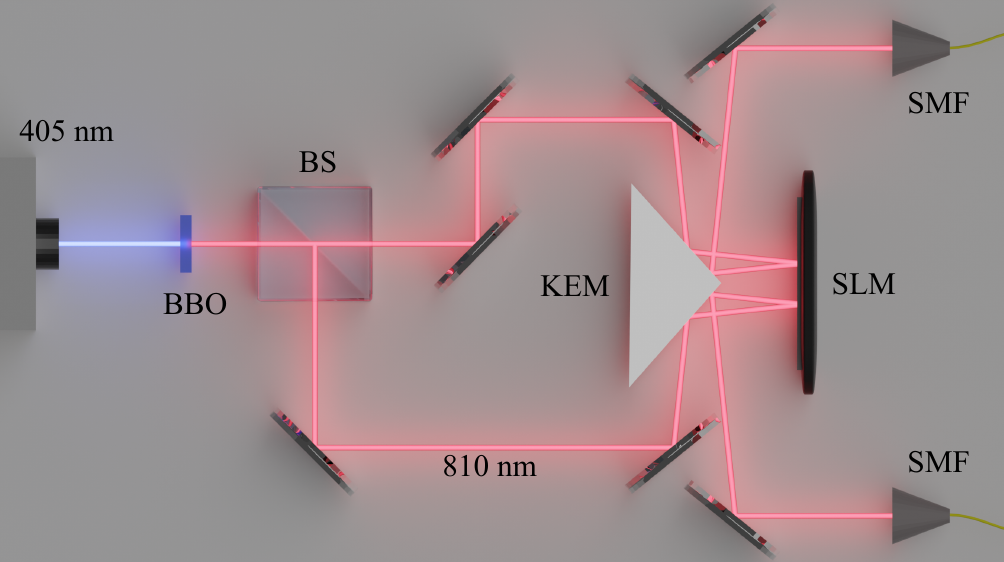}
\caption{\label{FIG:arreglo} Schematic sketch of our Type-I collinear SPDC source, with the pump beam at 405 nm. The experiment requires modulating both photons with one SLM. This was achieved with a Knife Edge Mirror (KEM).}
\end{figure}

To experimentally verify the tuning of the joint probability, we build a type-I collinear SPDC source. In Fig.~\ref{FIG:arreglo} we show a schematic sketch of our experimental arrangement, where a 405~nm CW laser beam was coupled into a single-mode fiber in order to obtain a clean Gaussian beam ($l_p=0$) as the pump beam. Then, the down-converted photons are filtered with a 500~nm long pass filter, followed by a band-pass filter centered at 810~nm, with a width of  $\Delta \lambda=$ 10~nm. The filtered photons are separated with a 50:50 Beam Splitter (BS) and then projected to the desired HIG states by using a Knife Edge Mirror (KEM) and single Spatial Light Modulator (SLM) with split screen in order to project the desired modes independently for each photon. The image of the crystal is mapped into the SLM with a telescope and finally a demagnifier telescope is used to reduce the incoming beam to the single mode fiber's core size (SMF). With the experimental technique presented in Ref.~\cite{bentley-2006}, we display the desired holograms of the HIG modes in the SLM, thus we are able to project the down converted photons into any specific HIG spatial mode. To measure the joint probabilities, we tune the ellipticity in a range from 0 to 10.

\subsection{Both photons with distinct helicity}

We start our experimental analysis by tuning the probability of finding the entangled state in the HIG modes basis, considering that each photon has different helicity, which is given by $|F_{p,m}|^2$. First, it is worth pointing out some characteristics of this probability. As we take greater $p$ and $m$ values, the modal probability decreases for values $\epsilon_s\neq \epsilon_i$. Then, as $\epsilon_s$ take values far from $\epsilon_i$, the coefficients $D^\sigma_{ln}(\epsilon_s)$ and $D^\sigma_{ln}(\epsilon_i)$ tend to be more and more distinct from each other, such as $D^\sigma_{ln}(\epsilon_s)<D^\sigma_{ln}(\epsilon_i)$, and their product is smaller than in the $\epsilon_s=\epsilon_i$ case. This behavior is stronger for greater values of $p$, as the terms of the sum on Eq.~(\ref{FCoef}) increase with this index. This effect is a characteristic of the HIG states, not related to the nature of the entangled state generated by the SPDC process, which was already discussed~\cite{krenn-2013}.

\begin{figure}[h]
\includegraphics[scale=0.35]{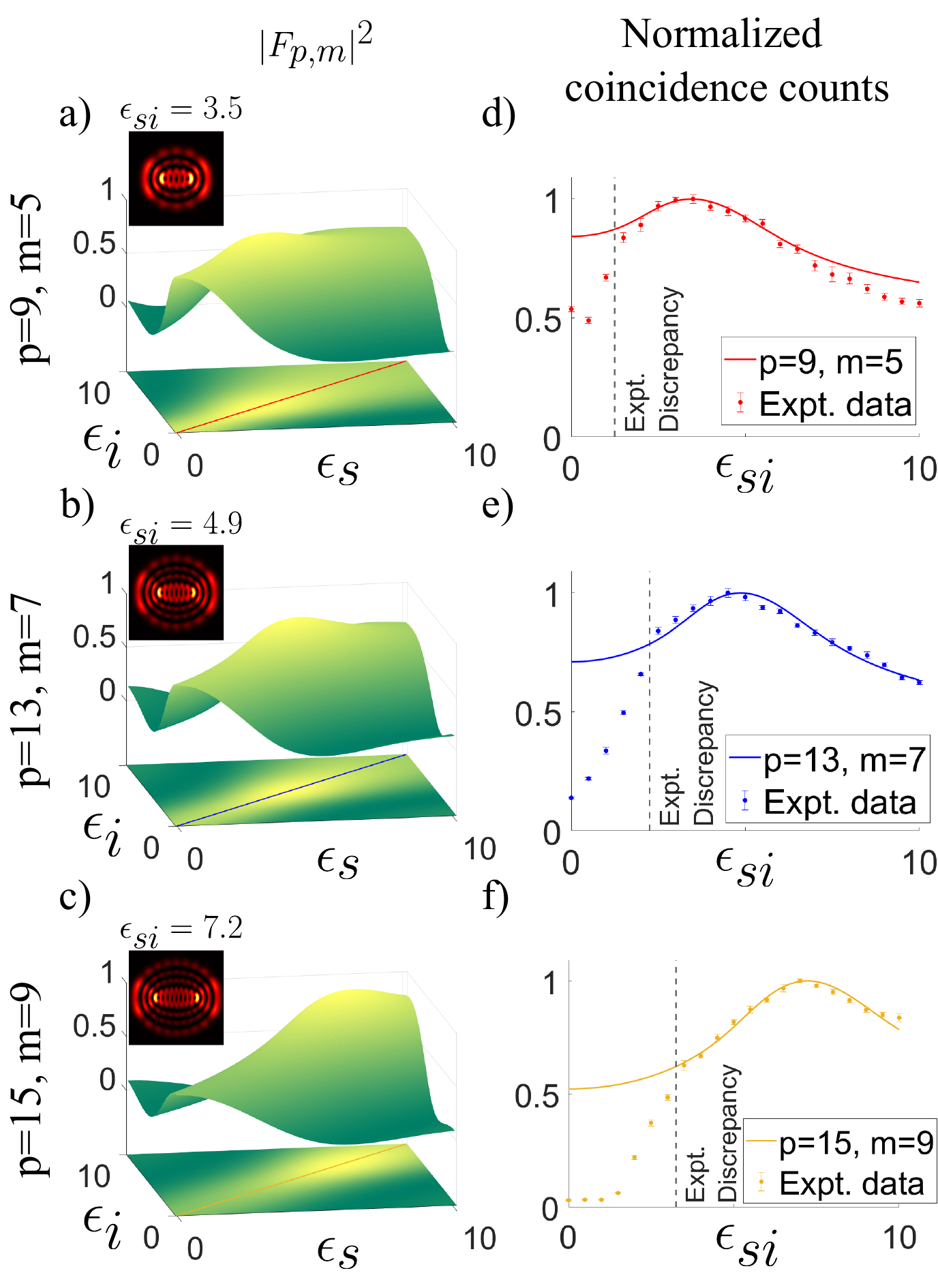}
\caption{\label{FIG:masmenos} Normalized probabilities $|F_{p,m}|^2$ for the HIG modes with a) $p=9,m=5$, b) $p=13,m=7$ and c) $p=15,m=9$. As higher orders are taken, the $\epsilon_{si}$ value that maximizes the joint probability moves to higher values. The insets shows the transverse intensity of each projected mode for the $\epsilon_{si}$ values that maximize the corresponding probability. The contour lines $\epsilon_s=\epsilon_i\equiv\epsilon_{si}$ are shown aside its corresponding probability function, together with the experimental data.}
\end{figure}

The second main characteristic of $|F_{p,m}|^2$, is that it reaches its maximum value when $\epsilon_s=\epsilon_i\equiv\epsilon_{si}$, having a global maximum in some specific $\epsilon_{si}$, which varies depending on the HIG mode. For modes with the lowest $m$ value (i.e. $m=1$ or $2$, according to the parity of the state), for a given $p$, the probability for $\epsilon_{si}$ monotonically decreases; this decay becomes more pronounced for higher $m$ modes. On the other hand, for modes with the highest $m$ value (i.e. $m=p$), for a given $p$, the probability for $\epsilon_{si}$ monotonically increases. For the cases when $m$ is between its minimum and maximum possible values, for some $p$ given, the $\epsilon_{si}$ value that maximizes the joint probability lies at a specific value between zero and infinity, this value increases along with the order of the mode. If the modal number $p$ is fixed, the $\epsilon_{si}$ value that maximizes the joint probability, increases along with $m$. On the contrary, if we fix the modal number $m$, the $\epsilon_{si}$ value that maximizes the joint probability decreases when the modal number $p$ increases. This behaviour is depicted in Fig.~\ref{FIG:masmenos} for some HIG modes, along with
corresponding experimental measurements.

\begin{figure}[H]
\includegraphics[scale=0.35]{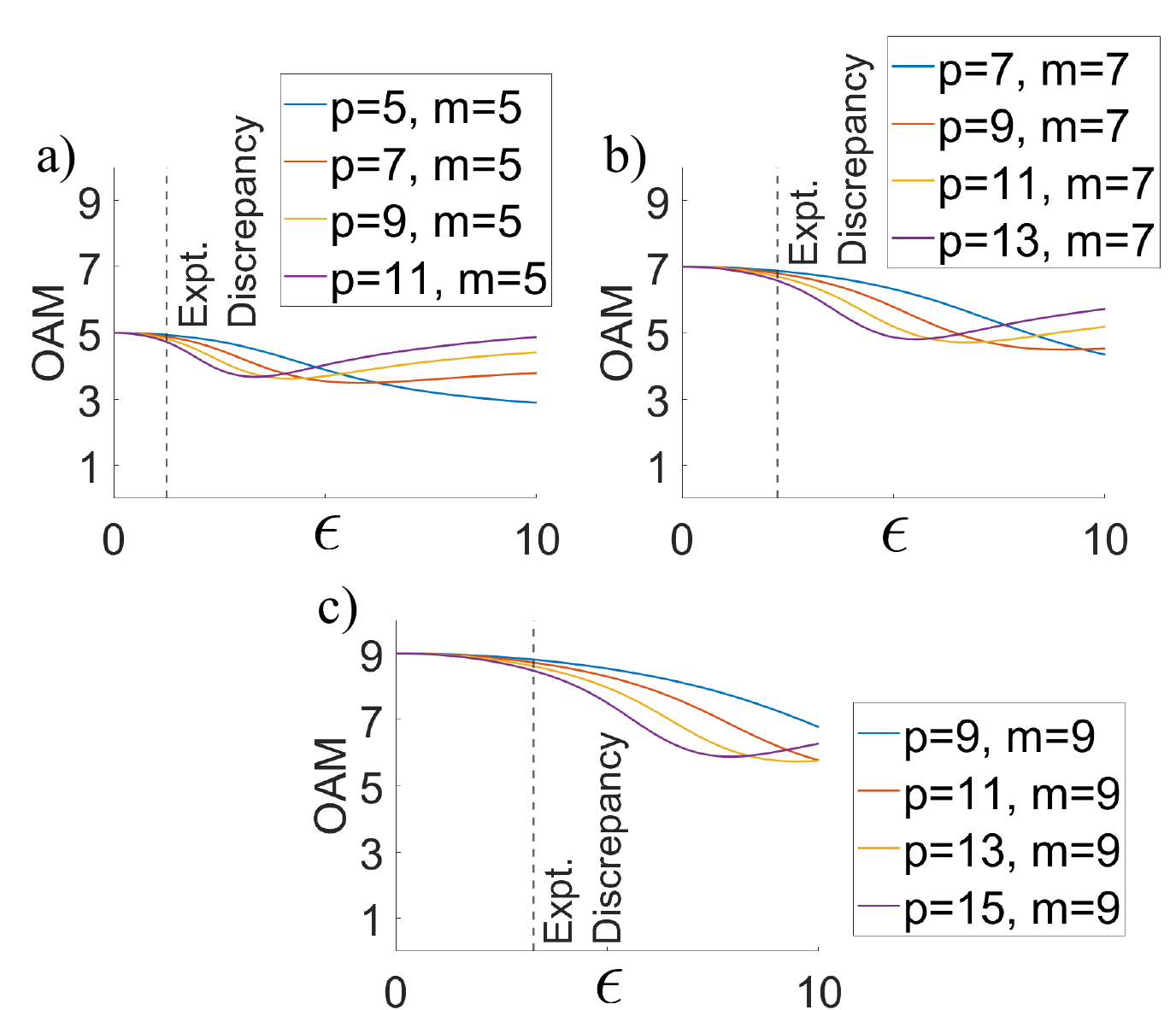}
\caption{\label{FIG:errores} Expectation value of OAM divided by $\hbar$ (Eq.~(\ref{OAMfraccionario})), for the HIG modes in a single-photon number state. The experimental discrepancy line is placed at the beginning of the ellipticity range in which the experimental and theoretical data agree, as in figures~\ref{FIG:masmenos} d), e) and f).}
\end{figure}

The tuning of probability from Fig.~\ref{FIG:masmenos} is related to the coincidence counts, when varying the ellipticity of the two holograms on which the photons are projected. The experimental contour lines are shown in Fig.~\ref{FIG:masmenos} d), e) and f). The measurements for low $\epsilon$ values does not match with the theoretical curves. A explanation for this is given by the description of the OAM content for the IG Fock states~(Eq.~(\ref{OAMfraccionario})). Let's take the example of measuring the state with the mode $p=13, m=7$. To measure a photon in a specific HIG mode, we transfer the opposite amount of OAM with the SLM, in order to couple it into a SMF. For this mode and low ellipticity values, we need to transfer an approximate OAM value of $7$, but there are other HIG modes that also carry such an OAM value for this $\epsilon$ values, this is shown in Fig. \ref{FIG:errores} b), from which the amount of OAM is nearly the same for different modes. Qualitatively, the interval where the experimental data does not match with the theoretical description ($\epsilon\in(0,2)$) is the same interval where the HIG Fock states share nearly the same amount of OAM. This happens also experimentally in the $p=9, m=5$, and in the $p=15, m=9$ case (Fig.~\ref{FIG:errores} a) and c)).

These discrepancies given by cross correlated terms could affect the use of the of Eq.~(\ref{SPDCIGdos}) in quantum communications, since it might compromise the security of the protocol. This behaviour can be improved by using modulated amplitude in the SLMs, as has been done before for other families of spatial light modes~\cite{zhang-2018}.

The behavior of the joint probability as an ellipticity function is attributed to the SPDC's OAM spectrum; as the $C_{n,n}^{l,-l}$ coefficients limit the products of the $D_{ln}(\epsilon_{si})$ coefficients in Eq.~(\ref{FCoef}), which endows this characteristic that prefers certain values $\epsilon_{si}$ with which the maximum probability is found. 

\subsection{Both photons with equal helicity}
\label{Both photons with equal helicity}

We just analyzed the tuning of the probability of finding both photons with distinct helicity, when the SPDC state is described in terms of the HIG modes. This case is a natural extension of the standard decomposition in the LG basis (Eq.~(\ref{estadoSPDC})). Now, we are interested in tuning the probability of finding both photons with the same helicity, in HIG Fock states, which is given by $|G_{p,m}|^2$. Once again there are some characteristics of this probability that need to be highlighted.

\begin{figure}[H]
\includegraphics[scale=0.35]{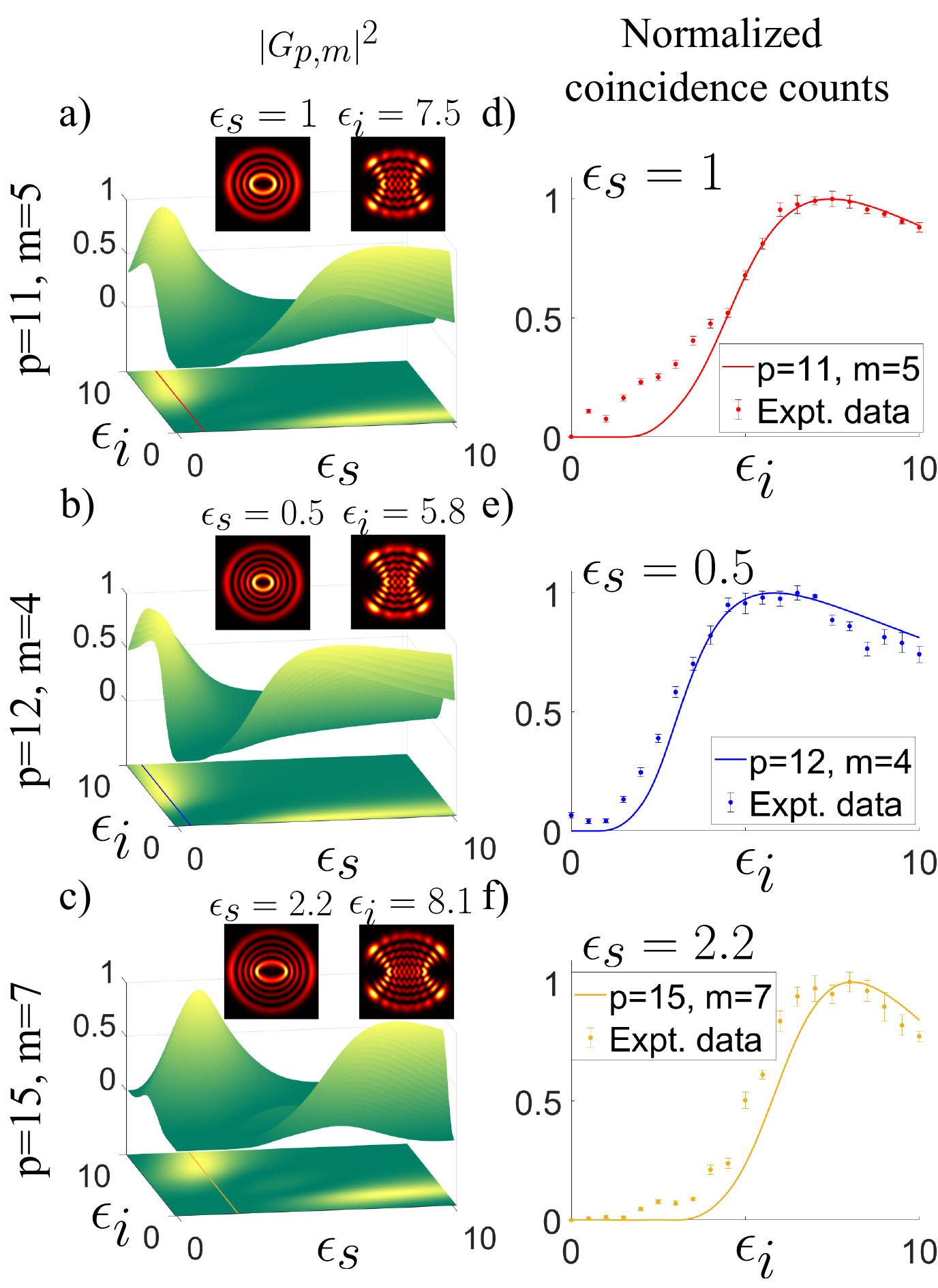}
\caption{\label{FIG:masmas} Normalized probabilities $|G_{p,m}|^2$ for the HIG modes with a) $p=11,m=5$, b) $p=12,m=4$ and c) $p=15,m=7$. As higher orders are taken, the values $\epsilon_{s}$ and $\epsilon_{i}$ that maximizes the joint probability moves to higher values. The insets shows the transverse intensity of each projected mode for the $\epsilon_{s}$ and $\epsilon_{i}$ values that maximize the corresponding probability. The contour lines are fixed at the $\epsilon_s$ value that maximizes the probability. These are shown aside its corresponding probability function, together with the experimental data.}
\end{figure}

If we fix the modal number $m$, the $\epsilon_{s}$ and $\epsilon_{i}$ values, that maximizes the joint probability, decreases when the modal number $p$ increases. On the contrary, if we fix the modal number $m$, the $\epsilon_{s}$ and $\epsilon_{i}$ values, that maximizes the joint probability decreases when the modal number $p$ increases. An example is depicted in Fig.~\ref{FIG:masmas}, along with experimental results.

\hfill

As it is shown, in the Helical Ince-Gauss basis the photon-pair entangled state does not behave as in the Laguerre-Gauss basis. In the LG case the expectation value of finding both photons with the same helicity in the state is null, but in the Helical Ince-Gauss basis the probability of finding both photons with the same helicity does not vanish. This is due that the expansion coefficients for the Helical Ince-Gauss beams differs for the different parities $D_{ln}^e \neq D_{ln}^o$, except in the limit $\epsilon\to 0$, where we recover the zero probability as described in the LG case.

\subsection{Bell test}
\label{section3}

Finally we seek to demonstrate that the SPDC state certainly is composed of the two HIG Bell states $\ket{\Psi^+}$ and $\ket{\Phi^+}$, as seen in Eq.~(\ref{SPDCIGdos}). The entanglement and steering of the component state $\ket{\Psi^+}$, has already been confirmed~\cite{krenn-2013}. However, since the contribution of the state $\ket{\Phi^+}$ has never been taken into account, we must verify the existence of such state. 

In order to confirm the existence of the component states and its entanglement we perform a Bell test to violate the CHSH-Bell inequality, which is a variant of Bell's inequality, in the same experimental manner as has been done for the other OAM basis~\cite{leach-2009, dada-2011, zhang-2018,moreau-2019, walborn-2005,mclaren-2012, krenn-2013, tang-2016}. This inequality is formulated for binary measurements, where the outcomes of the measurements are represented by binary values~\cite{clauser-1969}.

To do so, we measure the correlations of the signal and idler photons on a superposition state by defining a HIG Bloch sphere ~\cite{krenn-2013} where the poles of the sphere are HIG modes, and each point on the equator represents a specific superposition with a well-defined phase as be represented by,
\begin{equation}\label{proyectorinicial}
\ket{\theta}= \frac{1}{\sqrt{2}}\left(e^{i\theta}\ket{I^{+,\epsilon}_{p,m}}+e^{-i\theta}\ket{I^{-,\epsilon}_{p,m}}\right),
\end{equation}

\noindent so the angles of Eq.~(\ref{proyectorinicial}) works as the detector settings to calculate the Bell parameter~\cite{leach-2009},
\begin{equation}\label{primerS}
    S = |E(\theta_s,\theta_i) - E(\theta_s,\theta_i') + E(\theta_s',\theta_i) + E(\theta_s',\theta_i')|,
\end{equation}

\noindent where the correlation function is given by,
\begin{widetext}
\begin{equation}\label{E_iELL}
E(\theta_s,\theta_i)=\frac{C(\theta_s,\theta_i)+C(\theta_s+\pi/2,\theta_i+\pi/2)-C(\theta_s+\pi/2,\theta_i)-C(\theta_s,\theta_i+\pi/2)}{C(\theta_s,\theta_i)+C(\theta_s+\pi/2,\theta_i+\pi/2)+C(\theta_s+\pi/2,\theta_i)+C(\theta_s,\theta_i+\pi/2)}.
\end{equation}
\end{widetext}

If the value of $S$ is greater than $2$, then the Bell inequality is violated, which confirms the entanglement of the state.

The joint probability $C(\theta_s,\theta_i)$ is calculated with the desired entangled state. For instance, projecting the superposition states $\bra{\theta_s}$ and $\bra{\theta_i}$, on the state $\ket{\Psi^+}$, and squaring its absolute value,
\begin{equation}\label{coinc1}
C(\theta_s,\theta_i) 
= |\bra{\theta_s}\bra{\theta_i}\ket{\Psi^+}|^2= \cos^2(\theta_i-\theta_s),
\end{equation}

\noindent substituting this into $E(\theta_s,\theta_i)$ (Eq.~(\ref{E_iELL})), and then in $S$ (Eq.~(\ref{primerS})), we find that the set of angles that maximize $S$ to $2\sqrt{2}$ for the state $\ket{\Psi^+}$ are $\theta_s=0^\circ$, $\theta_i=22.5^\circ$, $\theta_s'=45^\circ$, $\theta_i'=67.5^\circ$.

Analogously, projecting the superposition states $\bra{\theta_s}$ and $\bra{\theta_i}$, on the state $\ket{\Phi^+}$, and squaring its absolute value,
\begin{equation}\label{coinc2}
C(\theta_s,\theta_i) 
= |\bra{\theta_s}\bra{\theta_i}\ket{\Phi^+}|^2= \cos^2(\theta_i+\theta_s),
\end{equation}

\noindent here we find that the set of angles that maximize $S$ to $2\sqrt{2}$ for the state $\ket{\Phi^+}$ are $\theta_s=90^\circ$, $\theta_i=22.5^\circ$, $\theta_s'=45^\circ$, $\theta_i'=67.5^\circ$.

It is noteworthy that the set of angles that maximize the Bell parameter for the state $\ket{\Psi^+}$ are the angles from where a minimum of $S$ is found for the state $\ket{\Phi^+}$, and vice versa. So by using the correct set of angles, it is possible to confirm to contribution and the entanglement of each component state. 

To calculate the Bell parameter $S$ for the photon-pair state, given by SPDC (Eq.~(\ref{estadoSPDC})), is convenient to write Eq.~(\ref{proyectorinicial}) in terms of the even and odd Laguerre-Gauss Fock states, 
\begin{equation}\label{pasos1}
\ket{\theta}= \cos(\theta)\sum_{l,n} D_{ln}^e(\epsilon)\ket{L^e_{l,n}}-\sin(\theta)\sum_{l,n} D_{ln}^o(\epsilon)\ket{L^o_{l,n}}.
\end{equation}

Then, the joint probability of finding the entangled state $|\bra{\theta_s}\bra{\theta_i}\ket{\Psi_{\textrm{SPDC}}}|^2$ is given by
\begin{gather}
C(\theta_s,\theta_i)=|2\cos(\theta_s)\cos(\theta_i)\sum_{l,n} C_{n,n}^{l,-l} D_{ln}^e(\epsilon_s)D_{ln}^e(\epsilon_i)\nonumber\\
+2\sin(\theta_s)\sin(\theta_i)\sum_{l,n} C_{n,n}^{l,-l}D_{ln}^o(\epsilon_s) D_{ln}^o (\epsilon_i)|^2.\label{coincSPDC}
  \end{gather}
  
As expected, when $\epsilon\to0$, this joint probability becomes $C(\theta_s,\theta_i)\propto \cos^2(\theta_i-\theta_s)$ as Eq.~(\ref{coinc1}).

\begin{figure}[H]
\includegraphics[scale=0.35]{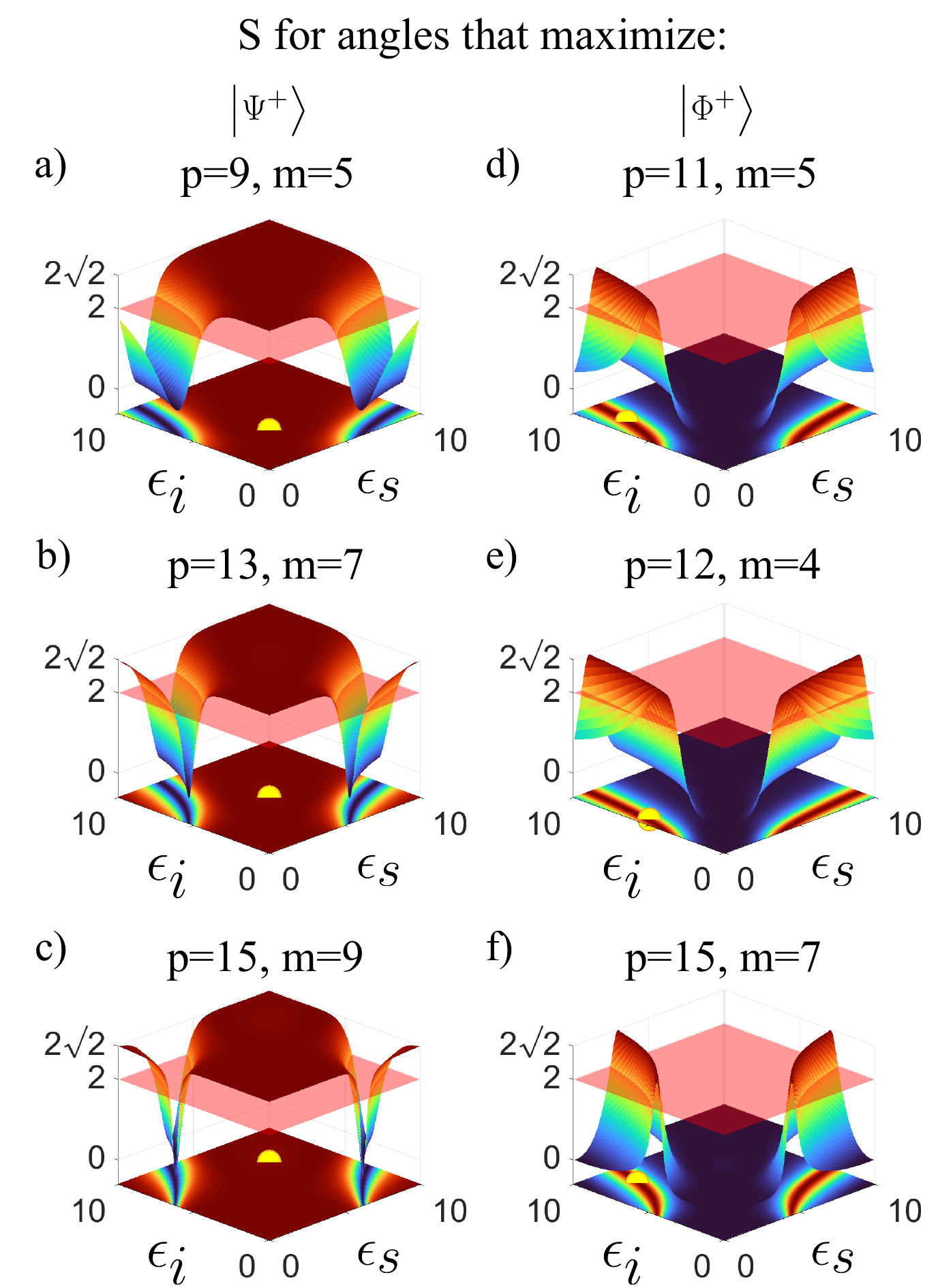}
\caption{\label{FIG:Bell} $S$ value as a function of $\epsilon_s$ and $\epsilon_s$ for the same HIG Fock states as in Figs.~\ref{FIG:masmenos} and~\ref{FIG:masmas}. The first column corresponds to $S$ calculated in angles that maximize the violation of the $\ket{\Psi^+}$ state. The second column corresponds to $S$ calculated in angles that maximize the violation of the $\ket{\Phi^+}$. The red transparent plane is placed in $S=2$. The yellow circle on each graph represents the $\epsilon_s$ and $\epsilon_i$ values in which the test was experimentally made.}
\end{figure}

By substituting Eq.~(\ref{coincSPDC}) in Eq.~(\ref{E_iELL}), in order to obtain the Bell parameter, we can graph $S$ as a function of $\epsilon_s$ and $\epsilon_i$. In Fig.~\ref{FIG:Bell} we fixed the set of angles that maximize the inequality for the states $\ket{\Psi^+}$ and $\ket{\Phi^+}$. As it is shown, most of the ellipticity values maximize the Bell parameter $S$ for the case where the entanglement is given by $\ket{\Psi^+}$, which corresponds to the same values that maximize the probability of finding such state. For the angles that maximizes the inequality for the state $\ket{\Phi^+}$, there are narrower regions of ellipticities for which the Bell inequality is violated. However, even if there are fewer values that maximize this case, this still verifies the form of the state given in Eq.~(\ref{SPDCIGdos}).  The yellow circles in Fig.~\ref{FIG:Bell} show the ellipticities in which the Bell parameter was experimentally calculated. These results are shown in tables~\ref{tab:BELL1} and~\ref{tab:BELL2}.
\break

\begin{table}[H]
\caption{\label{tab:BELL1}%
Experimental values of $S$, measured with angles that maximize the Bell parameter of $\ket{\Psi^+}$, for the same modes as in Fig.~\ref{FIG:masmenos}.
}
\begin{ruledtabular}
\begin{tabular}{lcc}
\textrm{Quantum numbers}&
\textrm{Expt. $S$}&
\textrm{Violation by $\sigma$}\\
\colrule
 $p=$9, $m=$5, $\epsilon_{si}$=3.5 & 2.56$\pm$0.06 & 9\\
 $p=$13, $m=$7, $\epsilon_{si}$=5.0 & 2.30$\pm$0.07 & 4\\
 $p=$15, $m=$9, $\epsilon_{si}$=7.0 & 2.33$\pm$0.09 &  4\\
\end{tabular}
\end{ruledtabular}
\end{table}
\begin{table}[H]
\caption{\label{tab:BELL2}%
Experimental values of $S$, measured with angles that maximize the Bell parameter of $\ket{\Phi^+}$, for the same modes as in Fig.~\ref{FIG:masmas}.
}
\begin{ruledtabular}
\begin{tabular}{lcc}
\textrm{Quantum numbers}&
\textrm{Expt. $S$}&
\textrm{Violation by $\sigma$}\\
\colrule
 $p=$11, $m=$5, $\epsilon_{s}$=1.0, $\epsilon_{i}$=7.5 & 2.43$\pm$0.07 & 6\\
 $p=$12, $m=$4, $\epsilon_{s}$=0.5, $\epsilon_{i}$=5.5 & 2.29$\pm$0.09 & 3\\
 $p=$15, $m=$7, $\epsilon_{s}$=2.2, $\epsilon_{i}$=8 & 2.21$\pm$0.08 &  3\\
\end{tabular}
\end{ruledtabular}
\end{table}


\section{Conclusions}
\label{section5}
We have provided a detailed theoretical-experimental description of the two-photon OAM entangled state, generated by a SPDC type-I collinear process, described in terms of the Helical Ince-Gauss modes basis. In this work we found two main results: 1) The two-photon OAM entangled state $\ket{\Psi_{\textrm{SPDC}}}$ is composed of a superposition of two symmetric Bell states, namely $\ket{\Psi^+}$ and $\ket{\Phi^+}$, whose probabilities can be independently selected by adjusting the ellipticity of the HIG modes basis in which the state is being projected. The contribution of each of these entangled states to the total SPDC state have been verified through the violation of CHSH-Bell inequalities ($S>2$). 2) It is possible to modulate the joint probability of detecting both down-converted photons also by changing the ellipticity parameter of the HIG modes used for state projection. This behavior has been proved by implementing detailed measurements of HIG modal joint probability.

\begin{acknowledgments}
This work was supported by Consejo Nacional de Ciencia y Tecnología, CONACYT Mexico (Grant Fronteras de la Ciencia No. 217559).
\end{acknowledgments}

\nocite{*}

%

\end{document}